\documentclass[twocolumn,english,aps,prl,superscriptaddress]{revtex4-1}
\usepackage[T1]{fontenc}
\usepackage[latin9]{inputenc}
\usepackage{color}
\usepackage{amsmath}
\usepackage{amssymb}
\usepackage{graphicx}
\usepackage{esint}

\makeatletter
\usepackage{babel}

\usepackage{babel}

\usepackage{babel}

\makeatother

\usepackage{babel}
\begin{document}

\title{Coupled Spin-Light dynamics in Cavity Optomagnonics}

\author{Silvia Viola Kusminskiy}

\affiliation{Institute for Theoretical Physics, University Erlangen-Nürnberg,
Staudtstraße 7, 91058 Erlangen, Germany}

\author{Hong X. Tang}

\affiliation{Department of Electrical Engineering, Yale University, New Haven,
Connecticut 06511, USA}

\author{Florian Marquardt}

\affiliation{Institute for Theoretical Physics, University Erlangen-Nürnberg,
Staudtstraße 7, 91058 Erlangen, Germany}

\affiliation{Max Planck Institute for the Science of Light, Günther-Scharowsky-Straße
1, 91058 Erlangen, Germany}
\begin{abstract}
Experiments during the past two years have shown strong resonant photon-magnon
coupling in microwave cavities, while coupling in the optical regime
was demonstrated very recently for the first time. Unlike with microwaves,
the coupling in optical cavities is parametric, akin to optomechanical
systems. This line of research promises to evolve into a new field
of optomagnonics, aimed at the coherent manipulation of elementary
magnetic excitations by optical means. In this work we derive the
microscopic optomagnonic Hamiltonian. In the linear regime the system
reduces to the well-known optomechanical case, with remarkably large
coupling. Going beyond that, we study the optically induced nonlinear
classical dynamics of a macrospin. In the fast cavity regime we obtain
an effective equation of motion for the spin and show that the light
field induces a dissipative term reminiscent of Gilbert damping. The
induced dissipation coefficient however can change sign on the Bloch
sphere, giving rise to self-sustained oscillations. When the full
dynamics of the system is considered, the system can enter a chaotic
regime by successive period doubling of the oscillations. 
\end{abstract}
\maketitle

\section{Introduction \label{sec:Introduction}}

The ability to manipulate magnetism has played historically an important
role in the development of information technologies, using the magnetization
of materials to encode information. Today's research focuses on controlling
individual spins and spin currents, as well as spin ensembles, with
the aim of incorporating these systems as part of quantum information
processing devices. \cite{Tserkovnyak2005,Chumak2015,Krawczyk2014,Kurizki2015}.
In particular the control of elementary excitations of magnetically
ordered systems \textendash denominated magnons or spin waves, is
highly desirable\textcolor{black}{{} since their frequency is broadly
tunable (ranging from MHz to THz) \cite{Stancil2009,Chumak2015} while
they can have very long lifetimes}, especially for insulating materials
like the ferrimagnet yttrium iron garnet (YIG) \cite{Serga2010}.
The collective character of the magnetic excitations moreover render
these robust against local perturbations.

Whereas the good magnetic properties of YIG have been known since
the 60s, it is only recently that coupling and controlling spin waves
with electromagnetic radiation in solid-state systems has started
to be explored. Pump-probe experiments have shown ultrafast magnetization
switching with light \cite{Stanciu2007,Kirilyuk2010,Lambert2015},
and strong photon-magnon coupling has been demonstrated in microwave
cavity experiments \cite{Huebl2013,Zhang2014a,Tabuchi2014,Goryachev2014,Bourhill2015,Haigh2015a,Bai2015a,Zhang2015b,Lambert2015a}
\textendash including the photon-mediated coupling between a superconducting
qubit and a magnon mode \cite{Tabuchi2015a}. Going beyond microwaves,
this points to the tantalizing possibility of realizing \emph{optomagnonics}:
the coupled dynamics of magnons and photons in the optical regime,
which can lead to coherent manipulation of magnons with light. The
coupling between magnons and photons in the optical regime differs
from that of the microwave regime, where resonant matching of frequencies
allows for a linear coupling: one magnon can be converted into a photon,
and viceversa \cite{Soykal2010,Cao2015,ZareRameshti2015}. In the
optical case instead, the coupling is a three-particle process. This
accounts for the frequency mismatch and is generally called parametric
coupling. The mechanism behind the optomagnonic coupling is the Faraday
effect, where the angle of polarization of the light changes as it
propagates through a magnetic material. Very recent first experiments
in this regime show that this is a promising route, by demonstrating
coupling between optical modes and magnons, and advances in this field
are expected to develop rapidly \cite{Zhang2015,Haigh2015,Zhang2015a,Hisatomi2016,Osada2016}.
\begin{figure}
\includegraphics[width=1\columnwidth]{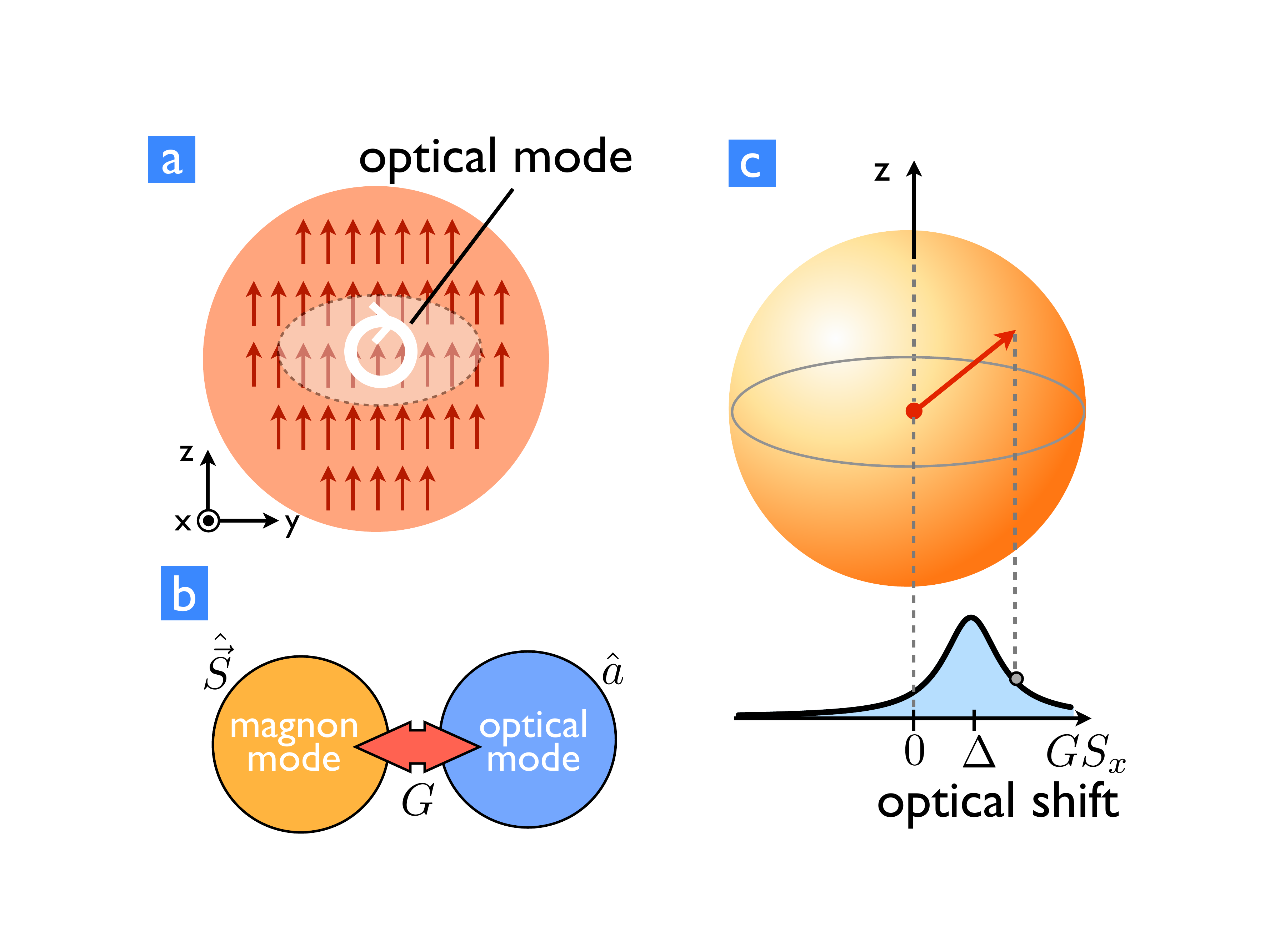} \caption{(Color online) Schematic configuration of the model considered. (a)
Optomagnonic cavity with homogeneous magnetization along the \emph{z}-axis
and a localized optical mode with circular polarization in the \emph{y-z}-plane.
(b) The homogeneous magnon mode couples to the optical mode with strength
$G$. (c) Representation of the magnon mode as a macroscopic spin
on the Bloch sphere, whose dynamics is controlled by the coupling
to the driven optical mode.}
\label{SetUp} 
\end{figure}

In this work we derive and analyze the basic optomagnonic Hamiltonian
that allows for the study of solid-state cavity optomagnonics. The
parametric optomagnonic coupling is reminiscent of optomechanical
models. In the magnetic case however, the relevant operator that couples
to the optical field is the spin, instead of the usual bosonic field
representing a mechanical degree of freedom. Whereas at small magnon
numbers the spin can be replaced by a harmonic oscillator and the
ideas of optomechanics \cite{AspelmeyerRMP14} carry over directly,
for general trajectories of the spin this is not possible. This gives
rise to rich non-linear dynamics which is the focus of the present
work. Parametric spin-photon coupling has been studied previously in atomic ensembles \cite{Hammerer2010,Brahms2010}. In this work we focus on solid-state systems with magnetic order and derive the corresponding optomagnonic Hamiltonian. After obtaining the general Hamiltonian, we consider
a simple model which consists of one optical mode coupled to a homogeneous
Kittel magnon mode \cite{Kittel1948}. We study the classical dynamics
of the magnetic degrees of freedom and find magnetization switching,
self-sustained oscillations, and chaos, tunable by the light field
intensity. 

The manuscript is ordered as follows. In Sec. \eqref{sec:Model} we
present the model and the optomagnonic Hamiltonian which is the basis
of our work. In Sec. \eqref{sec:Relation-to-optomechanics} we discuss
briefly the connection of the optomagnonic Hamiltonian derived in
this work and the one used in optomechanic systems. In Sec. \eqref{sec:Microscopic}
we derive the optomagnonic Hamiltonian from microscopics, and give
an expression for the optomagnonic coupling constant in term of material
constants. In Sec. \eqref{sec:Spin-dynamics} we derive the classical
coupled equations of motion of spin and light for a homogeneous magnon
mode, in which the spin degrees of freedom can be treated as a macrospin.
In Sec. \eqref{sec:Fast-cavity} we obtain the effective equation
of motion for the macrospin in the fast-cavity limit, and show the
system presents magnetization switching and self oscillations. We
treat the full (beyond the fast-cavity limit) optically induced nonlinear
dynamics of the macrospin in Sec. \eqref{sec:Full-nonlinear-dynamics},
and follow the route to chaotic dynamics. In Sec. \eqref{sec:Discussion}
we sketch a qualitative phase diagram of the system as a function
of coupling and light intensity, and discuss the experimental feasibility
of the different regimes. An outlook and conclusions are found in
Sec. \eqref{sec:Outlook}. In the Appendix we give details of some
of the calculations in the main text, present more examples of nonlinear
dynamics as a function of different tuning parameters, and compare
optomagnonic \emph{vs. }optomechanic attractors.

\section{Model\label{sec:Model}}

Further below, we derive the optomagnonic Hamiltonian which forms
the basis of our work: 
\begin{equation}
H=-\hbar\Delta\hat{a}^{\dagger}\hat{a}-\hbar\Omega\hat{S}_{z}+\hbar G\hat{S}_{x}\hat{a}^{\dagger}\hat{a}\,,\label{eq:Hamiltonian}
\end{equation}
where $\hat{a}^{\dagger}$ ($\hat{a}$) is the creation (annihilation)
operator for a cavity mode photon. We work in a frame rotating at
the laser frequency $\omega_{las}$, and $\Delta=\omega_{las}-\omega_{cav}$
is the detuning with respect to the optical cavity frequency $\omega_{cav}$.
Eq.~\eqref{eq:Hamiltonian} assumes a magnetically ordered system
with (dimensionless) macrospin ${\bf S}=(S_{x},S_{y},S_{z})$ with
magnetization axis along $\mathbf{\hat{z}}$, and a precession frequency
$\Omega$ which can be controlled by an external magnetic field \footnote{note that this frequency however depends on the magnetic field \emph{inside}
the sample, and hence it depends on its geometry and the corresponding
demagnetization fields.}. The coupling between the optical field and the spin is given by
the last term in Eq.~\eqref{eq:Hamiltonian}, \textcolor{black}{where
we assumed (see below) that light couples only to the $x-$} component
of the spin as shown in Fig.~\eqref{SetUp}. The coefficient $G$
denotes the parametric optomagnonic coupling. We will derive it in
terms of the Faraday rotation, which is a material-dependent constant.

\subsection{Relation to optomechanics \label{sec:Relation-to-optomechanics}}

Close to the ground state, for deviations such that $\delta S\ll S$
(with $S=|\mathbf{S}|$), we can treat the spin in the usual way as
a harmonic oscillator, $\hat{S}_{x}\approx\sqrt{S/2}(\hat{b}+\hat{b}^{\dagger})$,
with $\left[\hat{b},\hat{b}^{\dagger}\right]=1$. Then the optomagnonic
interaction $\hbar G\hat{S}_{x}\hat{a}^{\dagger}\hat{a}\approx\hbar G\sqrt{S/2}\hat{a}^{\dagger}\hat{a}(\hat{b}+\hat{b}^{\dagger})$
becomes formally equivalent to the well-known opto\emph{mechanical}
interaction \cite{AspelmeyerRMP14}, with bare coupling constant $g_{0}=G\sqrt{S/2}$.
All the phenomena of optomechanics apply, including the ``optical
spring'' (here: light-induced changes of the magnon precession frequency)
and optomagnonic cooling at a rate $\Gamma_{{\rm {\rm opt}}}$, and
the formulas (as reviewed in Ref. \cite{AspelmeyerRMP14}) can be
taken over directly. All these effects depend on the light-enhanced
coupling $g=g_{0}\alpha$, where $\alpha=\sqrt{n_{{\rm phot}}}$ is
the cavity light amplitude. For example, in the sideband-resolved
regime ($\kappa\ll\Omega$, where $\kappa$ is the optical cavity
decay rate) one would have $\Gamma_{{\rm opt}}=4g^{2}/\kappa$. If
$g>\kappa$, one enters the strong-coupling regime, where the magnon
mode and the optical mode hybridize and where coherent state transfer
is possible. A Hamiltonian of the form of Eq.~(\ref{eq:Hamiltonian})
is also encountered for light-matter interaction in atomic ensembles
\cite{Hammerer2010}, and its explicit connection to optomechanics
in this case was discussed previously in Ref. \cite{Brahms2010}.
In contrast to such non-interacting spin ensembles, the confined magnon
mode assumed here can be frequency-separated from other magnon modes.

\subsection{Microscopic magneto-optical coupling $G$\label{sec:Microscopic}}

In this section we derive the Hamiltonian presented in Eq. \eqref{eq:Hamiltonian}
starting from the microscopic magneto-optical effect in Faraday-active
materials. The Faraday effect is captured by an effective permittivity
tensor that depends on the magnetization $\mathbf{M}$ in the sample.
We restrict our analysis to non-dispersive isotropic media and linear
response in the magnetization, and relegate magnetic linear birefringence
effects which are quadratic in $\mathbf{M}$ (denominated the Cotton-Mouton
or Voigt effect) for future work \cite{Landau1984,Stancil2009}. In
this case, the permittivity tensor acquires an antisymmetric imaginary
component and can be written as $\mbox{\ensuremath{\varepsilon_{ij}\left(\mathbf{M}\right)}=\ensuremath{\varepsilon_{0}}\ensuremath{\left(\varepsilon\delta_{ij}-if\sum_{k}\epsilon_{ijk}M_{k}\right)}}$,
where $\varepsilon_{0}$ ($\varepsilon$) is the vacuum (relative)
permittivity, $\epsilon_{ijk}$ the Levi-Civita tensor and $f$ a
material-dependent constant \cite{Landau1984} (here and in what follows,
Latin indices indicate spatial components). The Faraday rotation per
unit length 
\begin{equation}
\theta_{F}=\frac{\omega fM_{s}}{2c\sqrt{\varepsilon}}\,,\label{eq:thetaF}
\end{equation}
depends on the frequency $\omega$, the vacuum speed of light $c$,
and the saturation magnetization $M_{s}$. The magneto-optical coupling
is derived from the time-averaged energy $\bar{U}=\frac{1}{4}\int{\rm d}\mathbf{r}\,\sum_{ij}\,E_{i}^{*}(\mathbf{r},t)\varepsilon_{ij}E_{j}(\mathbf{r},t)$,
using the complex representation of the electric field, $\left(\mathbf{E}+\mathbf{E^{*}}\right)/2$.
Note that $\bar{U}$ is real since $\varepsilon_{ij}$ is hermitean
\cite{Landau1984,Stancil2009}. The magneto-optical contribution is
\begin{equation}
\bar{U}_{MO}=-\frac{i}{4}\varepsilon_{0}f\int{\rm d}\mathbf{r}\,\mathbf{M}(\mathbf{r)}\cdot[\mathbf{E^{*}}\left(\mathbf{r}\right)\times\mathbf{E}\left(\mathbf{r}\right)]\,.\label{eq:UMO-1}
\end{equation}
This couples the magnetization to the spin angular momentum density
of the light field. Quantization of this expression leads to the optomagnonic
coupling Hamiltonian. A similar Hamiltonian is obtained in atomic
ensemble systems when considering the electric dipolar interaction
between the light field and multilevel atoms, where the spin degree
of freedom (associated with $\mathbf{M}(\mathbf{r})$ in our case)
is represented by the atomic hyperfine structure \cite{Hammerer2010}.
The exact form of the optomagnonic Hamiltonian will depend on the
magnon and optical modes. In photonic crystals, it has been demonstrated
that optical modes can be engineered by nanostructure patterning \cite{Joannopoulos2008},
and magnonic-crystals design is a matter of intense current research
\cite{Krawczyk2014}. The electric field is easily quantized, $\mathbf{\hat{E}^{(+)}}(\mathbf{r},t)=\sum_{\beta}\mathbf{E}_{\beta}(\mathbf{r})\hat{a}_{\beta}(t)$,
where $\mathbf{E}_{\beta}(\mathbf{r})$ indicates the $\beta^{{\rm th}}$
eigenmode of the electric field (eigenmodes are indicated with Greek
letters in what follows). The magnetization requires more careful
consideration, since $\mathbf{M}(\mathbf{r})$ depends on the local
spin operator which, in general, cannot be written as a linear combination
of bosonic modes. There are however two simple cases: (i) small deviations
of the spins, for which the Holstein-Primakoff representation is linear
in the bosonic magnon operators, and (ii) a homogeneous Kittel mode
$\mathbf{M}(\mathbf{r})=\mathbf{M}$ with macrospin $\mathbf{S}$.
In the following we treat the homogeneous case, to capture nonlinear
dynamics. From Eq.~\eqref{eq:UMO-1} we then obtain the coupling
Hamiltonian $\hat{H}_{MO}=\hbar\sum_{j\beta\gamma}\hat{S}_{j}G_{\beta\gamma}^{j}\hat{a}_{\beta}^{\dagger}\hat{a}_{\gamma}$
with 
\begin{equation}
G_{\beta\gamma}^{j}=-i\frac{\varepsilon_{0}f\,M_{s}}{4\hbar S}\sum_{mn}\epsilon_{jmn}\int{\rm d}\mathbf{r}E_{\beta m}^{*}(\mathbf{r)}E_{\gamma n}(\mathbf{r})\,,\label{eq:G_jbg-1}
\end{equation}
where we replaced $M_{j}/M_{s}=\hat{S}_{j}/S$, with $S$ the extensive
total spin (scaling like the mode volume). One can diagonalize the
hermitean matrices $G^{j}$, though generically not simultaneously.
In the present work, we treat the conceptually simplest case of a
strictly diagonal coupling to some optical eigenmodes ($G_{\beta\beta}^{j}\neq0$
but $G_{\alpha\beta}^{j}=0$). This is precluded only if the optical
modes are both time-reversal invariant (${\bf E}_{\beta}$ real-valued)
and non-degenerate. In all the other cases, a basis can be found in
which this is valid. For example, a strong static Faraday effect will
turn optical circular polarization modes into eigenmodes. Alternatively,
degeneracy between linearly polarized modes implies we can choose
a circular basis.

Consider circular polarization (R/L) in the $y-z$-plane, such that
$G^{x}$ is diagonal while $G^{y}=G^{z}=0$. Then we find

\begin{equation}
G_{LL}^{x}=-G_{RR}^{x}=G=\frac{1}{S}\frac{c\,\theta_{F}}{4\sqrt{\varepsilon}}\xi\,,\label{eq:GMO}
\end{equation}
where we used Eq.~\eqref{eq:thetaF} to express the coupling via
the Faraday rotation $\theta_{F}$, and where $\xi$ is a dimensionless
overlap factor that reduces to $1$ if we are dealing with plane waves
(see App.~\ref{sec:SuppG}). Thus, we obtain the coupling Hamiltonian
$H_{MO}=\hbar G\hat{S}_{x}(\hat{a}_{L}^{\dagger}\hat{a}_{L}-\hat{a}_{R}^{\dagger}\hat{a}_{R})$.
This reduces to Eq.~(\ref{eq:Hamiltonian}) if the incoming laser
drives only one of the two circular polarizations.

The coupling $G$ gives the \emph{magnon precession} frequency shift
\emph{per} photon. It decreases for larger magnon mode volume, in
contrast to $GS$, which describes the overall \emph{optical} shift
for saturated spin ($S_{x}=S$). For YIG, with $\varepsilon\approx5$
and $\theta_{F}\approx200^{{\rm o}}{\rm cm}^{-1}$ \cite{Weber1994,Stancil2009},
we obtain $GS\approx10^{10}{\rm Hz}$ (taking $\xi=1$), which can
easily become comparable to the precession frequency $\Omega$. The
ultimate limit for the magnon mode volume is set by the optical wavelength,
$\sim(1\mu m)^{3}$, which yields $S\sim10^{10}$. Therefore $G\approx1{\rm Hz}$,
whereas the coupling to a single magnon would be remarkably large:
$g_{0}=G\sqrt{S/2}\approx0.1{\rm MHz}$. This provides a strong incentive
for designing small magnetic structures, by analogy to the scaling
of piezoelectrical resonators \cite{Fan2015}. Conversely, for a macroscopic
volume of $(1mm)^{3}$, with $S\sim10^{19}$, this reduces to $G\approx10^{-9}{\rm Hz}$
and $g_{0}\approx10{\rm Hz}$.

\section{Spin dynamics\label{sec:Spin-dynamics}}

The coupled Heisenberg equations of motion are obtained from the Hamiltonian
in Eq. \eqref{eq:Hamiltonian} by using $\left[\hat{a},\hat{a}^{\dagger}\right]=1$,
$\left[\hat{S}_{i},\hat{S}_{j}\right]=i\epsilon_{ijk}\hat{S}_{k}$.
We next focus on the classical limit, where we replace the operators
by their expectation values: 
\begin{eqnarray}
\dot{a} & = & -i\left(GS_{x}-\Delta\right)a-\frac{\kappa}{2}\left(a-\alpha_{{\rm max}}\right)\nonumber \\
\dot{\mathbf{S}} & = & \left(Ga^{*}a\mathbf{\,e}_{x}-\Omega\,\mathbf{e}_{z}\right)\times\mathbf{S}+\frac{\eta_{{\rm G}}}{S}(\mathbf{\dot{S}}\times\mathbf{S})\,.\label{eq:EOM}
\end{eqnarray}
Here we introduced the laser amplitude $\alpha_{{\rm max}}$ and the
intrinsic spin Gilbert-damping \cite{Gilbert2004}, characterized
by $\eta_{{\rm G}}$, due to phonons and defects ($\eta_{{\rm G}}\approx10^{-4}$
for YIG \footnote{In the magnetic literature, $\eta_{{\rm G}}$ is denoted as $\alpha$
\cite{Stancil2009}. }). After rescaling the fields (see App..~\ref{sec:SuppLinearEOM}),
we see that the classical dynamics is controlled by only five dimensionless
parameters: $\frac{GS}{\Omega},\,\frac{G\alpha_{{\rm max}}^{2}}{\Omega},\,\frac{\Delta}{\Omega},\,\frac{\kappa}{\Omega},\,\eta_{{\rm G}}$.
These are independent of $\hbar$ as expected for classical dynamics.

In the following we study the nonlinear classical dynamics of the
spin, and in particular we treat cases where the spin can take values
on the whole Bloch sphere and therefore differs significantly from
a harmonic oscillator, deviating from the optomechanics paradigm valid
for $\delta S\ll S$. The optically induced tilt of the spin can be
estimated from Eq. \eqref{eq:EOM} as $\delta S/S=G|a|^{2}/\Omega\sim G\alpha_{{\rm max}}^{2}/\Omega=B_{{\rm \alpha_{{\rm max}}}}/\Omega$,
where $B_{\alpha_{{\rm max}}}=G\alpha_{{\rm max}}^{2}$ is an optically
induced effective magnetic field. We would expect therefore unique
optomagnonic behavior (beyond optomechanics) for large enough
light intensities, such that $B_{{\rm \alpha_{{\rm max}}}}$ is of
the order of or larger than the precession frequency $\Omega$. We
will show however that, in the case of blue detuning, even small light
intensity can destabilize the original magnetic equilibrium of the
uncoupled system, provided the intrinsic Gilbert damping is small.
\begin{figure}[b]
\includegraphics[width=1\columnwidth]{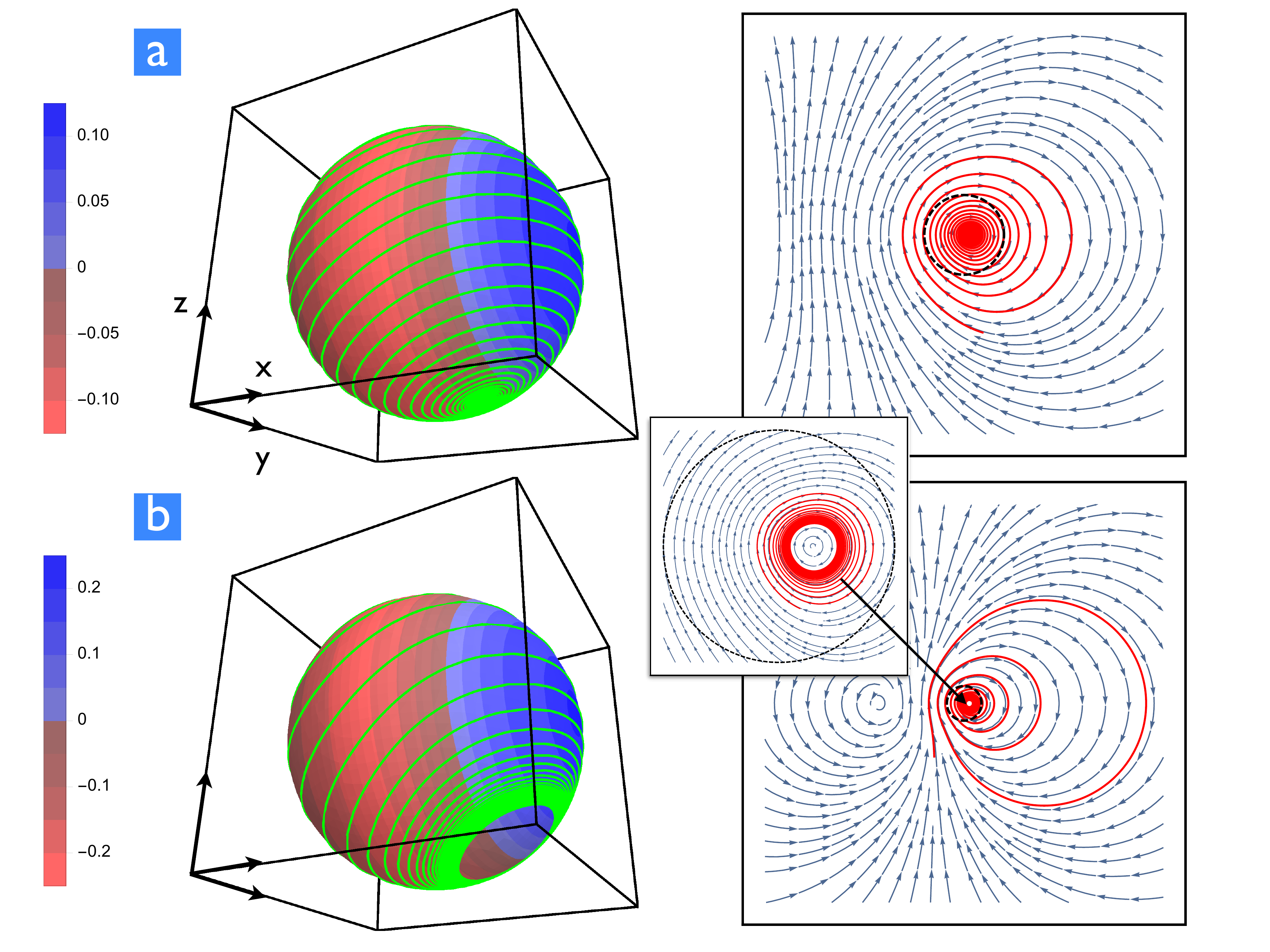} \caption{(Color online) Spin dynamics (fast cavity limit) at blue detuning
$\Delta=\Omega$ and fixed $GS/\Omega=2$, $\kappa/\Omega=5$, $\eta_{{\rm G}}=0$.
The left column depicts the trajectory (green full line) of a spin
(initially pointing near the north pole) on the Bloch sphere. The
color scale indicates the optical damping $\eta_{{\rm opt}}$. The
right column shows a stereographic projection of the spin's trajectory
(red full line). The black dotted line indicates the equator (invariant
under the mapping), while the north pole is mapped to infinity. The
stream lines of the spin flow are also depicted (blue arrows). (a)
Magnetization switching behavior for light intensity $G\alpha_{{\rm max}}^{2}/\Omega=0.36$.
(b) Limit cycle attractor for larger light intensity $G\alpha_{{\rm max}}^{2}/\Omega=0.64$.}
\label{AdDyn} 
\end{figure}

\subsection{Fast\emph{ }cavity\emph{ }regime\label{sec:Fast-cavity}}

As a first step we study a spin which is slow compared to the cavity,
where $G\dot{S}_{x}\ll\kappa^{2}$. In that case we can expand the
field $a(t)$ in powers of $\dot{S}_{x}$ and obtain an effective
equation of motion for the spin by integrating out the light field.
We write $a(t)=a_{0}(t)+a_{1}(t)+\ldots$, where the subscript indicates
the order in $\dot{S}_{x}$. From the equation for $a(t)$, we find
that $a_{0}$ fulfills the instantaneous equilibrium condition 
\begin{equation}
a_{0}(t)=\frac{\kappa}{2}\alpha_{{\rm max}}\frac{1}{\frac{\kappa}{2}-i\left(\Delta-GS_{x}(t)\right)}\,,\label{eq:a0}
\end{equation}
from which we obtain the correction $a_{1}$: 
\begin{equation}
a_{1}(t)=-\frac{1}{\frac{\kappa}{2}-i\left(\Delta-GS_{x}\right)}\frac{\partial a_{0}}{\partial S_{x}}\dot{S}_{x}\,.\label{eq:a1}
\end{equation}
To derive the effective equation of motion for the spin, we replace
$|a|^{2}\approx|a_{0}|^{2}+a_{1}^{*}a_{0}+a_{0}^{*}a_{1}$ in Eq.~\eqref{eq:EOM}
which leads to 
\begin{equation}
\dot{\mathbf{S}}=\mathbf{B}_{{\rm eff}}\times\mathbf{S}+\frac{\eta_{{\rm opt}}}{S}(\dot{S}_{x}\,\mathbf{e}_{x}\times\mathbf{S})+\frac{\eta_{{\rm G}}}{S}(\mathbf{\dot{S}}\times\mathbf{S})\,.\label{eq:EOM_eff}
\end{equation}
Here $\mathbf{B}_{{\rm eff}}=-\Omega\mathbf{e}_{z}+\mathbf{B}_{{\rm opt}}$,
where $\mathbf{B}_{{\rm opt}}(S_{x})=G|a_{0}|^{2}\,\mathbf{e}_{x}$
acts as an optically induced magnetic field. The second term is reminiscent
of Gilbert damping, but with spin-velocity component only along $\mathbf{e}_{x}$.
Both the induced field $\mathbf{B}_{{\rm opt}}$ and dissipation coefficient
$\eta_{{\rm opt}}$ depend explicitly on the instantaneous value of
$S_{x}(t)$: 
\begin{eqnarray}
\mathbf{B}_{{\rm opt}} & = & \frac{G}{[(\frac{\kappa}{2})^{2}+(\Delta-GS_{x})^{2}]}\left(\frac{\kappa}{2}\alpha_{{\rm max}}\right)^{2}\mathbf{e}_{x}\label{eq:B_ind}\\
\eta_{{\rm opt}} & = & -2G\kappa S\,|\mathbf{B}_{{\rm opt}}|\,\frac{(\Delta-GS_{x})}{[(\frac{\kappa}{2})^{2}+(\Delta-GS_{x})^{2}]^{2}}\,.\label{eq:G_ind}
\end{eqnarray}
This completes the microscopic derivation of the optical Landau-Lifshitz-Gilbert
equation for the spin, an important tool to analyze effective spin
dynamics in different contexts~\cite{Tserkovnyak2002}. We consider
the nonlinear adiabatic dynamics of the spin governed by Eq. \eqref{eq:EOM_eff}
below. Two distinct solutions can be found: generation of new stable
fixed points (magnetic switching) and optomagnonic limit cycles (self
oscillations).

Given our Hamiltonian (Eq.~\eqref{eq:Hamiltonian}), the north pole
is stable in the absence of optomagnonic coupling \textendash{} the
selection of this state is ensured by the intrinsic damping $\eta_{{\rm G}}>0$.
By driving the system this can change due to the energy pumped to
(or absorbed from) the spin, and the new equilibrium is determined
by $\mathbf{B}_{{\rm eff}}$ and $\eta_{{\rm opt}}$, when $\eta_{{\rm opt}}$
dominates over $\eta_{{\rm G}}$. Magnetic switching refers to the
rotation of the macroscopic magnetization by $\sim\pi$, to a new
fixed point near the south pole in our model. This can be obtained
for blue detuning $\Delta>0$, in which case $\eta_{{\rm opt}}$ is
negative either on the whole Bloch sphere (when $\Delta>GS$) or on
a certain region, as shown in Fig.~\eqref{AdDyn}a. Similar results
were obtained in the case of spin optodynamics for cold atoms systems~\cite{Brahms2010}.
The possibility of switching the magnetization direction in a controlled
way is of great interest for information processing with magnetic
memory devices, in which magnetic domains serve as information bits
\cite{Stanciu2007,Kirilyuk2010,Lambert2015}. Remarkably, we find
that for blue detuning, magnetic switching can be achieved for arbitrary
small light intensities in the case of $\eta_{{\rm G}}=0$. This is
due to runaway solutions near the north pole for $\Delta>0$, as discussed
in detail in App. \ref{sub:SuppSwitching}. In physical systems, the
threshold of light intensity for magnetization switching will be determined
by the extrinsic dissipation channels. 

For higher intensities of the light field, limit cycle attractors
can be found for $|\Delta|<GS$, where the optically induced dissipation
$\eta_{{\rm opt}}$ can change sign on the Bloch sphere (Fig.~\eqref{AdDyn}b).
The combination of strong nonlinearity and a dissipative term which
changes sign, leads the system into self sustained oscillations. The
crossover between fixed point solutions and limit cycle attractors
is determined by a balance between the detuning and the light intensity,
as discussed in App. \ref{sub:SuppSwitching}. Limit cycle attractors
require $B_{{\rm \alpha_{{\rm max}}}}/\Omega>|\Delta|/GS$ (note that
from \eqref{eq:G_ind} $B_{{\rm opt}}\sim B_{\alpha_{{\rm max}}}$
if $\kappa\gg(\Delta-GS)$ ).

We note that for both examples shown in Fig.~\eqref{AdDyn}, for
the chosen parameters we have $\eta_{{\rm opt}}\gg\eta_{{\rm G}}$
in the case of YIG, and hence taking $\eta_{{\rm G}}=0$ is a very
good approximation. More generally, from Eqs. \eqref{eq:B_ind} we
estimate $\eta_{opt}\sim GSB_{{\rm opt}}/\kappa^{3}$ and therefore
we can safely neglect $\eta_{{\rm G}}$ for $(\alpha_{_{{\rm max}}}G)^{2}S\gg\eta_{{\rm G}}\kappa^{3}$.
The qualitative results (limit cycle, switching) survive up to $\eta_{opt}\gtrsim\eta_{G}$,
although quantitatively modified as $\eta_{{\rm G}}$ is increased:
for example, the size of the limit cycle would change, and there would
be a threshold intensity for switching.

\subsection{Full nonlinear dynamics \label{sec:Full-nonlinear-dynamics}}

The nonlinear system of Eq.~\eqref{eq:EOM} presents even richer
behavior when we leave the fast cavity regime. For limit cycles near
the north pole, when $\delta S\ll S$, the spin is well approximated
by a harmonic oscillator, and the dynamics is governed by the attractor
diagram established for optomechanics \cite{Marquardt2006}. In contrast,
larger limit cycles will display novel features unique to optomagnonics,
on which we focus here.

\begin{figure}
\includegraphics[width=1\columnwidth]{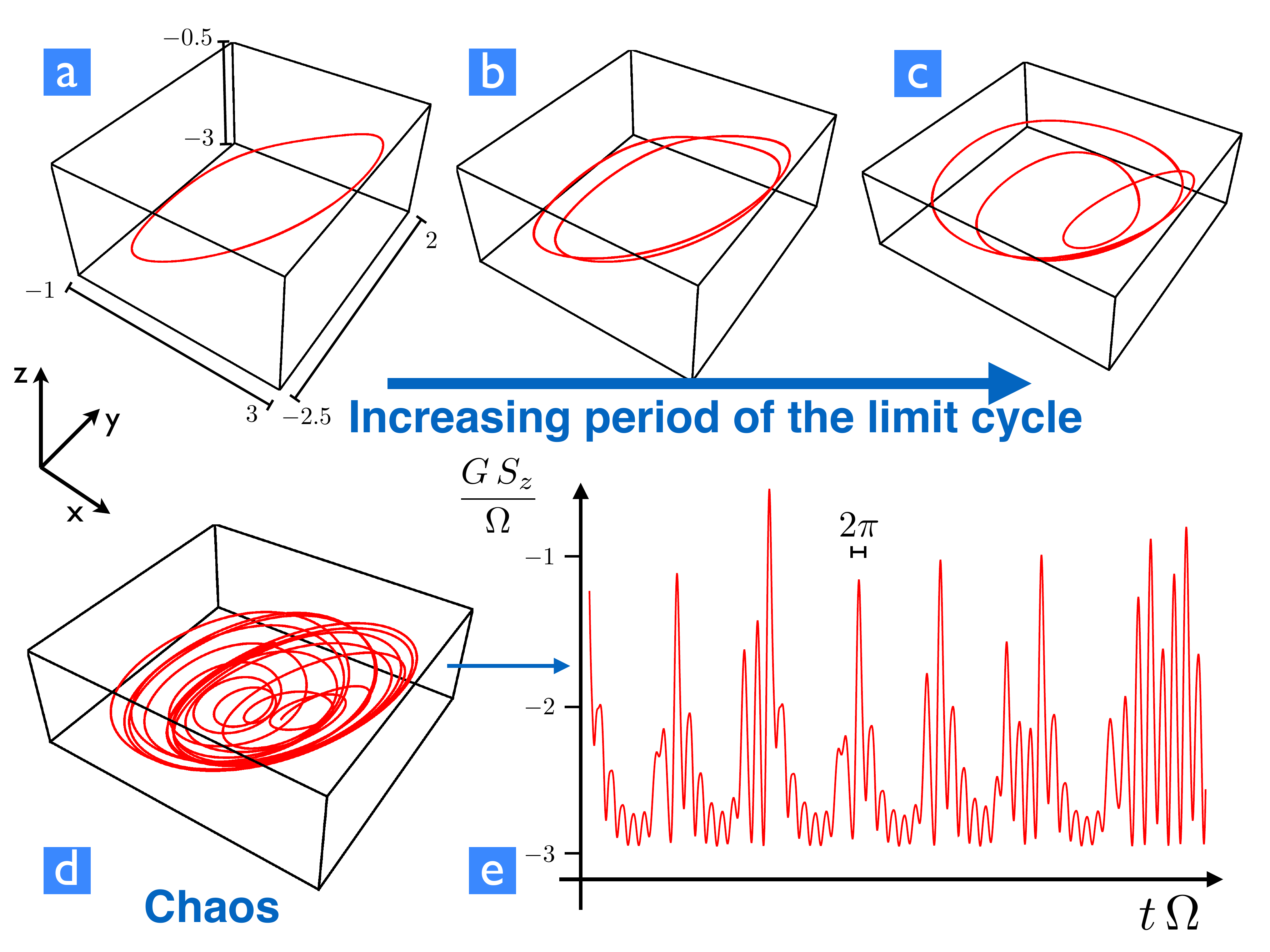} \caption{(Color Online) Full non-linear spin dynamics and route to chaos for
$GS/\Omega=3$ and $G\alpha_{{\rm max}}^{2}/\Omega=1$ ($\eta_{{\rm G}}=0$).
The system is blue detuned by $\Delta=\Omega$ and the dynamics, after
a transient, takes place in the southern hemisphere. The solid red
curves represent the spin trajectory after the initial transient,
on the Bloch sphere for (a) $\kappa/\Omega=3$, (b) $\kappa/\Omega=2$,
(c) $\kappa/\Omega=0.9$ , (d) $\kappa/\Omega=0.5$. (f) $S_{z}$
projection as a function of time for the chaotic case $\kappa/\Omega=0.5$.}
\label{FullDynFig} 
\end{figure}

\begin{figure}
\includegraphics[width=1\columnwidth]{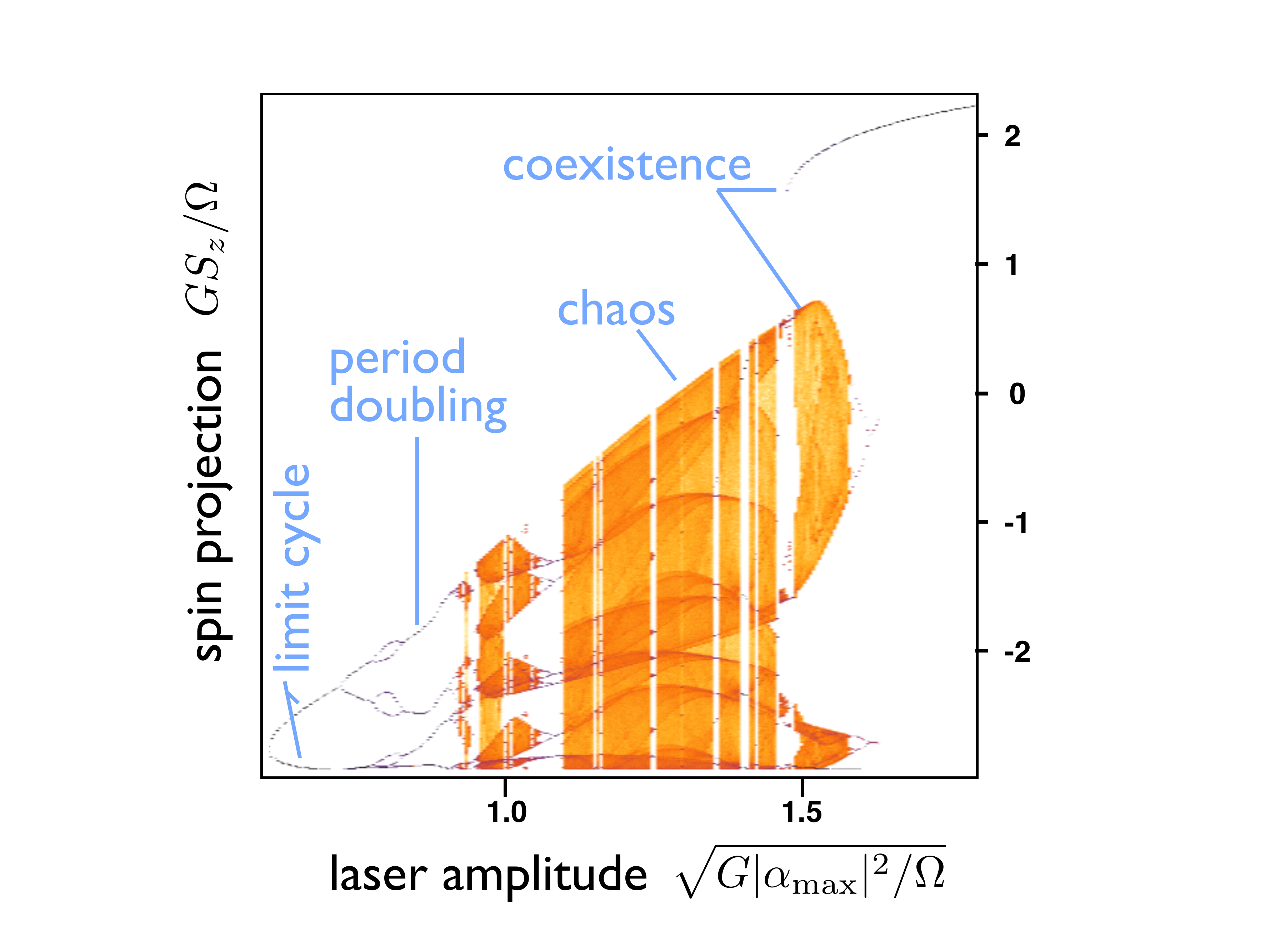} \caption{Bifurcation density plot for $GS/\Omega=3$ and $\kappa/\Omega=1$
at $\Delta=\Omega$ ($\eta_{{\rm G}}=0$), as a function of light
intensity. We plot the $S_{z}$ values attained at the turning points
($\dot{S}{}_{z}=0$). For other possible choices (\emph{eg. $\dot{S}{}_{x}=0$}
) the overall shape of the bifurcation diagram is changed, but the
bifurcations and chaotic regimes remain at the same light intensities.
For the plot, 30 different random initial conditions were taken. }
\label{Bif} 
\end{figure}

Beyond the fast cavity limit, we can no longer give analytical expressions for the optically induced magnetic field and dissipation. Moreover, we can not define a coefficient $\eta_{\rm{opt}}$ since an expansion in $\dot{S_x}$ is not justified. We therefore resort to numerical analysis of the dynamics. Fig.~\eqref{FullDynFig} shows a route to chaos by successive period
doubling, upon decreasing the cavity decay $\kappa$. This route can
be followed in detail as a function of any selected parameter by plotting
the respective bifurcation diagram. This is depicted in Fig. \eqref{Bif}.
The plot shows the evolution of the attractors of the system as the
light intensity is increased. The figure shows the creation and expansion
of a limit cycle from a fixed point near the south pole, followed
by successive period doubling events and finally entering into a chaotic
region. At high intensities, a limit cycle can coexist with a chaotic
attractor. For even bigger light intensities, the chaotic attractor
disappears and the system precesses around the $\mathbf{\mathbf{e}_{x}}$
axis, as a consequence of the strong optically induced magnetic field.
Similar bifurcation diagrams are obtained by varying either $GS/\Omega$
or the detuning $\Delta/\Omega$ (see App. \ref{sub:SuppNonlinear}).

\section{Discussion\label{sec:Discussion}}

\begin{figure}
\includegraphics[width=1\columnwidth]{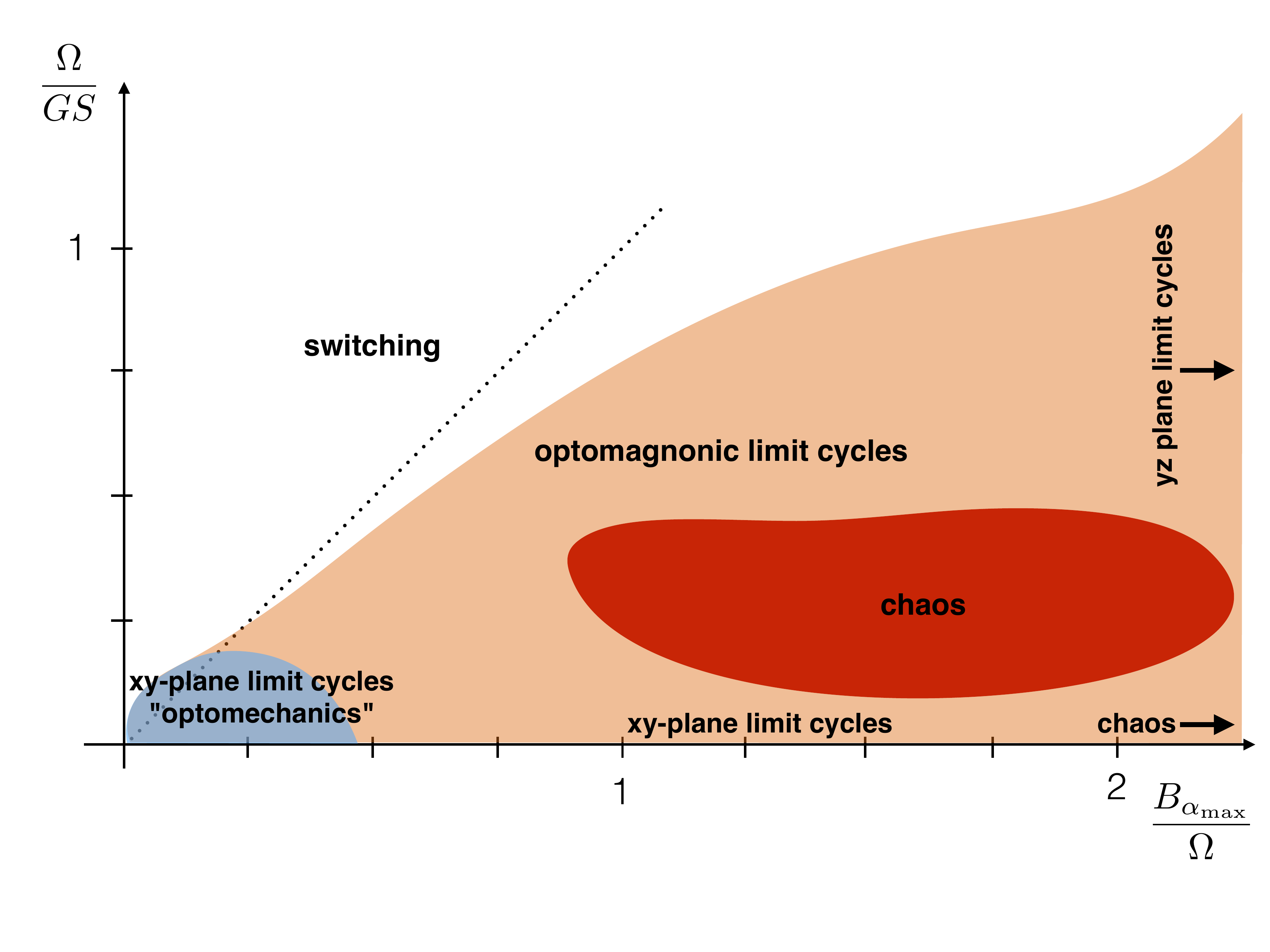}\caption{Phase diagram for blue detuning with $\Delta=\Omega$, as a function
of the inverse coupling strength $\Omega/GS$ and the optically induced
field $B_{\alpha_{{\rm max}}}/\Omega=G\alpha_{{\rm max}}^{2}/\Omega$.
Boundaries are qualitative. Switching, in white, refers to a fixed
point solution with the spin pointing near the south pole. Limit cycles
in the $xy$ plane are shaded in blue, and they follow the optomechanical
attractor diagram discussed in Ref. \cite{Marquardt2006}. For higher
$B_{\alpha_{{\rm max}}}$, chaos can ensue. Orange denotes the parameter
space in which limit cycles deviate markedly from optomechanical predictions.
These are not in the $xy$ plane and also undergo period doubling
leading to chaos. In red is depicted the area where pockets of chaos
can be found. For large $B_{\alpha_{{\rm max}}}/\Omega$, the limit
cycles are in the $yz$ plane. In the case of red detuning $\Delta=-\Omega$,
the phase diagram remains as is, except that instead of switching
there is a fixed point near the north pole.}
\label{Fig:PhaseBlue}
\end{figure}

We can now construct a qualitative phase diagram for our system. Specifically,
we have explored the qualitative behavior (fixed points, limit cycles,
chaos etc.) as a function of optomagnonic coupling and light intensity.
These parameters can be conveniently rescaled to make them dimensionless.
We chose to consider the ratio of magnon precession frequency to coupling,
in the form $\Omega/GS$. Furthermore, we express the light intensity
via the maximal optically induced magnetic field $B_{\alpha_{{\rm max}}}=G\alpha_{max}^{2}$.
The dimensionless coupling strength, once the material of choice is
fixed, can be tuned \emph{via} an external magnetic field which controls
the precession frequency $\Omega$. The light intensity can be controlled
by the laser.

We start by considering blue detuning, this is shown in Fig. \eqref{Fig:PhaseBlue}.
The ``phase diagram'' is drawn for $\Delta=\Omega$, and we set
$\kappa=\Omega$ and $\eta_{{\rm G}}=0$. We note that some of the
transitions are rather crossovers (``optomechanical limit cycles''
\emph{vs.} ``optomagnonic limit cycles''). In addition, the other
``phase boundaries'' are only approximate, obtained from direct
inspection of numerical simulations. These are not intended to be
exact, and are qualitatively valid for departures of the set parameters,
if not too drastic -- for example, increasing $\kappa$ will lead
eventually to the disappearance of the chaotic region.

As the diagram shows, there is a large range of parameters that lead
to magnetic switching, depicted in white. This area is approximately
bounded by the condition $B_{\alpha_{{\rm max}}}/\Omega\lesssim\Delta/GS$,
which in Fig. \eqref{Fig:PhaseBlue} corresponds to the diagonal since
we took $\Delta=\Omega$. This condition is approximate since it was
derived in the fast cavity regime, see App. \ref{sub:SuppSwitching}.
As discussed in Sec. \ref{sec:Spin-dynamics}, magnetic switching
should be observable in experiments even for small light intensity
in the case of blue detuning, provided that all non-optical dissipation
channels are small. The caveat of low intensity is a slow switching
time. For $B_{\alpha_{{\rm max}}}/\Omega\gtrsim\Delta/GS$, the system
can go into self oscillations and even chaos. For optically induced
fields much smaller than the external magnetic field, $B_{\alpha_{{\rm max}}}\ll\Omega$
we expect trajectories of the spin in the $xy$ plane, precessing
around the external magnetic field along ${\bf e_{z}}$ and therefore
the spin dynamics (after a transient) is effectively two-dimensional.
This is depicted by the blue-shaded area in Fig.\eqref{Fig:PhaseBlue}.
These limit cycles are governed by the optomechanical attractor diagram
presented in Ref. \cite{Marquardt2006}, as we show in App. \ref{sub:SuppOptomech}.
There is large parameter region in which the optomagnonic limit cycles deviate from the optomechanical
attractors. This is marked by orange in Fig.\eqref{Fig:PhaseBlue}.
As the light intensity is increased, for $\Omega/GS\ll1$ the limit
cycles remain approximately confined to the $xy$ plane but exhibit
deviations from optomechanics. This approximate confinement of the
trajectories to the $xy$ plane at large $B_{\alpha_{{\rm max}}}/\Omega$
($B_{\alpha_{{\rm max}}}/\Omega\gtrsim0.5$ for $\Delta=\Omega$)
can be understood qualitatively by looking at the expression of the
induced magnetic field $B_{{\rm opt}}$ deduced in the fast cavity
limit, Eq. \eqref{eq:B_ind}. Since we consider $\Delta=\Omega$,
$\Omega/GS\ll1$ implies $GS\gg\Delta$. In this limit, $B_{{\rm opt}}/\Omega$
can become very small and the spin precession is around the $\mathbf{e}_{z}$
axis. For moderate $B_{\alpha_{{\rm max}}}/\Omega$ and $\Omega/GS$,
the limit cycles are tilted and precessing around an axis determined
by the effective magnetic field, a combination of the optical induced
field and the external magnetic field. Blue detuning causes these
limit cycles to occur in the southern hemisphere. Period doubling
leads eventually to chaos. The region where pockets of chaos can be
found is represented by red in the phase diagram. For large light
intensity, such that $B_{\alpha_{{\rm max}}}\gg\Omega$, the optical
field dominates and the effective magnetic field is essentially along
the $\mathbf{e_{x}}$ axis. The limit cycle is a precession of the
spin around this axis.

According to our results optomagnonic chaos is attained for values
of the dimensionless coupling $GS/\Omega\sim1-10$ and light intensities
$G\alpha_{{\rm max}}^{2}/\Omega\sim0.1-1$. This implies a number
of circulating photons similar to the number of locked spins in the
material, which scales with the cavity volume. This therefore imposes
a condition on the minimum circulating photon density in the cavity.
For YIG with characteristic frequencies $\Omega\sim1-10{\rm GHz}$,
the condition on the coupling is easily fulfilled (remember $GS=10{\rm GHz}$
as calculated above). However the condition on the light intensity
implies a circulating photon density of $\sim10^{8}-10^{9}$ photons/${\rm \mu m^{3}}$
which is outside of the current experimental capabilities, limited
by the power a typical microcavity can support (around $\sim10^{5}$
photons/${\rm \mu m^{3}}$). On the other hand, magnetic switching
and self-sustained oscillations of the optomechanical type (but taking
place in the southern hemisphere) can be attained for low powers,
assuming all external dissipation channels are kept small. While self-sustained
oscillations and switching can be reached in the fast-cavity regime,
more complex nonlinear behavior such as period doubling and chaos
requires approaching sideband resolution. For YIG the examples in
Figs.~\ref{FullDynFig},~\ref{Bif} correspond to a precession frequency
$\Omega\approx3\cdot10^{9}{\rm Hz}$ (App. \ref{sub:SuppNonlinear}),
whereas $\kappa$ can be estimated to be $\sim10^{10}{\rm Hz}$, taking
into account the light absorption factor for YIG ($\sim0.3{\rm cm}^{-1}$)
\cite{Weber1994}.

For red detuning $\Delta<0$, the regions in the phase diagram remain
the same, except that instead of magnetic switching, the solutions
in this parameter range are fixed points near the north pole. This
can be seen by the symmetry of the problem: exchanging $\Delta\rightarrow-\Delta$
together with $\mathbf{e}_{x}\rightarrow-\mathbf{e}_{x}$ and $\mathbf{e}_{z}\rightarrow-\mathbf{e}_{z}$
leaves the problem unchanged. The limit cycles and trajectories follow
also this symmetry, and in particular the limit cycles in the $xy$
plane remain invariant.

\section{Outlook\label{sec:Outlook}}

The observation of the spin dynamics predicted here will be a sensitive
probe of the basic cavity optomagnonic model, beyond the linear regime.
Our analysis of the optomagnonic nonlinear Gilbert damping could be
generalized to more advanced settings, leading to optomagnonic reservoir
engineering (e.g. two optical modes connected by a magnon transition).
Although the nonlinear dynamics presented here requires light intensities
outside of the current experimental capabilities for YIG, it should
be kept in mind that our model is the simplest case for which highly
non-linear phenomena is present. Increasing the model complexity,
for example by allowing for multiple-mode coupling, could result in
a decreased light intensity requirement. Materials with a higher Faraday
constant would be also beneficial. In this work we focused on the
homogeneous Kittel mode. It will be an interesting challenge to study
the coupling to magnon modes at finite wavevector, responsible for
magnon-induced dissipation and nonlinearities under specific conditions~\cite{Clogston1956,Suhl1957,Gibson1984}.
The limit cycle oscillations can be seen as ``optomagnonic lasing'',
analogous to the functioning principle of a laser where energy is
pumped and the system settles in a steady state with a characteristic
frequency, and also discussed in the context of mechanics (``cantilaser''
\cite{Bargatin2003}). These oscillations could serve as a novel source
of traveling spin waves in suitable geometries, and the synchronization
of such oscillators might be employed to improve their frequency stability.
We may see the design of optomagnonic crystals and investigation of
optomagnonic polaritons in arrays. In addition, future cavity optomagnonics
experiments will allow to address the completely novel regime of cavity-assisted
coherent optical manipulation of nonlinear magnetic textures, like
domain walls, vortices or skyrmions, or even nonlinear spatiotemporal
light-magnon patterns. In the quantum regime, prime future opportunities
will be the conversion of magnons to photons or phonons, the entanglement
between these subsystems, and their applications to quantum communication
and sensitive measurements.

We note that different aspects of optomagnonic systems have been investigated
in a related work done simultaneously \cite{Liu2016}. Our work was
supported by an ERC-StG OPTOMECH and ITN cQOM. H.T. acknowledges support
by the Defense Advanced Research Projects Agency (DARPA) Microsystems
Technology Office/Mesodynamic Architectures program (N66001-11-1-4114) and an Air Force Office of Scientific Research (AFOSR) Multidisciplinary University Research Initiative grant (FA9550-15-1-0029). 

\bibliography{Optomagnonics}

\begin{widetext}
\appendix

\section{Optomagnonic coupling $G$ for plane waves}

\label{sec:SuppG} In this section we calculate explicitly the optomagnonic
coupling presented in Eq.. \eqref{eq:GMO} for the case of plane waves
mode functions for the electric field. We choose for definiteness
the magnetization axis along the $\mathbf{\hat{z}}$ axis, and consider
the case $G_{x\beta\gamma}\neq0$. The Hamiltonian $H_{MO}$ is then
diagonal in the the basis of circularly polarized waves, $\mathbf{e}_{R/L}=\frac{1}{\sqrt{2}}\left(\mathbf{e}_{y}\mp i\mathbf{e}_{z}\right)$.
The rationale behind choosing the coupling direction \emph{perpendicular}
to the magnetization axis, is to maximize the coupling to the magnon
mode, that is to the\emph{ deviations} of the magnetization with respect
to the magnetization axis. The relevant spin operator is therefore
$\hat{S}_{x}$, which represents the flipping of a spin. In the case
of plane waves, we quantize the electric field according to $\mathbf{\hat{E}^{+(-)}}(\mathbf{r},t)=+(-)i\sum_{_{j}}\mathbf{e}_{j}\sqrt{\frac{\hbar\omega_{j}}{2\varepsilon_{0}\varepsilon V}}\hat{a}_{j}^{(\dagger)}(t)e^{+(-)i\mathbf{k_{j}\cdot r}}\,,$
where $V$ is the volume of the cavity, $\mathbf{k}_{j}$ the wave
vector of mode $j$ and we have identified the positive and negative
frequency components of the field as $\mathbf{E}\rightarrow\hat{\mathbf{E}}^{+}$,
$\mathbf{E^{*}}\rightarrow\hat{\mathbf{E}}^{-}$. The factor of $\varepsilon_{0}\varepsilon$
in the denominator ensures the normalization $\hbar\omega_{j}=\varepsilon_{0}\varepsilon\langle j|\int d^{3}\mathbf{r}|\mathbf{E}(\mathbf{r})|^{2}|j\rangle-\varepsilon_{0}\varepsilon\langle0|\int d^{3}\mathbf{r}|\mathbf{E}(\mathbf{r})|^{2}|0\rangle$,
which corresponds to the energy of a photon in state $|j\rangle$
above the vacuum $|0\rangle$. For two degenerate (R/L) modes at frequency
$\omega$, using Eq. \eqref{eq:thetaF} we see that the frequency
dependence cancels out and we obtain the simple form for the optomagnonic
Hamiltonian $H_{MO}=\hbar G\hat{S}_{x}(\hat{a}_{L}^{\dagger}\hat{a}_{L}-\hat{a}_{R}^{\dagger}\hat{a}_{R})$
with $G=\frac{1}{S}\frac{c\,\theta_{F}}{4\sqrt{\varepsilon}}$. Therefore
the overlap factor $\xi=1$ in this case.

\section{Rescaled fields and linearized dynamics}

\label{sec:SuppLinearEOM}

To analyze Eq.~\eqref{eq:EOM} it is convenient to re-scale the fields
such that $a=\alpha_{{\rm max}}a'$, ${\bf S}=S{\bf S}'$ and measure
all times and frequencies in $\Omega$. We obtain the rescaled equations
of motion (time-derivatives are now with respect to $t'=\Omega t$)

\begin{eqnarray}
\dot{a}' & = & -i(\frac{GS}{\Omega}S_{x}'-\frac{\Delta}{\Omega})a'-\frac{\kappa}{2\Omega}(a'-1)\label{eq:REOM}\\
\dot{{\bf S}}' & = & \left(\frac{G\alpha_{{\rm max}}^{2}}{\Omega}|a'|^{2}{\bf e}_{x}-{\bf e}_{z}\right)\times{\bf S}'+\frac{\eta_{{\rm {G}}}}{S}\left(\dot{\mathbf{S}'}\times\mathbf{S}'\right)
\end{eqnarray}

If we linearize the spin-dynamics (around the north-pole, e.g.), we
should recover the optomechanics behavior. In this section we ignore
the intrinsic Gilbert damping term. We set approximately ${\bf S}'\approx(S'_{x},S'_{y},1)^{T}$
and from Eq.~\eqref{eq:REOM} we obtain

\begin{eqnarray}
\dot{S}_{x}' & = & S'_{y}\\
\dot{S}_{y}' & =- & \frac{G\alpha_{{\rm max}}^{2}}{\Omega}|a'|^{2}-S'_{x}
\end{eqnarray}

We can now choose to rescale further, via $S'_{x}=\left(\alpha_{{\rm max}}/\sqrt{S}\right)S''_{x}$
and likewise for $S_{y}'$. We obtain the following spin-linearized
equations of motion:

\begin{eqnarray}
\dot{S}_{x}'' & = & S''_{y}\\
\dot{S}_{y}'' & = & -\frac{G\sqrt{S}\alpha_{{\rm max}}}{\Omega}|a'|^{2}-S''_{x}\\
\dot{a}' & = & -i(\frac{G\sqrt{S}\alpha_{{\rm max}}}{\Omega}S_{x}''-\frac{\Delta}{\Omega})a'-\frac{\kappa}{2\Omega}(a'-1)
\end{eqnarray}

This means that the number of dimensionless parameters has been reduced
by one, since the two parameters initially involving G, $S$, and
$\alpha_{{\rm max}}$ have all been combined into

\begin{equation}
\frac{G\sqrt{S}\alpha_{{\rm max}}}{\Omega}
\end{equation}

In other words, for $S'_{x,y}=S_{x,y}/S\ll1$, the dynamics should
only depend on this combination, consistent with the optomechanical
analogy valid in this regime as discussed in the main text (where
we argued based on the Hamiltonian).

\section{Switching in the fast cavity limit }

\label{sub:SuppSwitching}From Eq. \eqref{eq:EOM_eff} in the weak dissipation
limit ($\eta_{G}\,\ll1$) we obtain 
\begin{align*}
\dot{S}_{x}= & \Omega S_{y}\\
\dot{S}_{y}= & -S_{z}B_{{\rm opt}}-\Omega S_{x}-\frac{\eta_{{\rm opt}}}{S}\dot{S}_{x}S_{z}\,,
\end{align*}
from where we obtain an equation of motion for $S_{x}$. We are interested
in studying the stability of the north pole once the driving is turned
on. Hence we set $S_{z}=S$, 
\[
\ddot{S}_{x}=-\Omega SB_{{\rm opt}}-\Omega^{2}S_{x}-\eta_{{\rm opt}}\Omega\dot{S}_{x}\,,
\]
and we consider small deviations $\delta S_{x}$ of $S_{x}$ from
the equilibrium position that satisfies $S_{x}^{0}=-SB_{{\rm opt}}/\Omega$,
where $B_{{\rm opt}}$ is evaluated at $S_{x}^{0}$. To linear order
we obtain 
\[
\ddot{\delta S}_{x}=-\Omega\left(\Omega+S\frac{\partial B_{{\rm opt}}}{\partial S_{x}}\right)\delta S_{x}+2GS\kappa\Omega B_{{\rm opt}}\frac{(\Delta+GSB_{{\rm opt}}/\Omega)}{\left[\left(\kappa/\Omega\right)^{2}+(\Delta+GSB_{{\rm opt}}/\Omega)^{2}\right]^{2}}\dot{\delta S}_{x}\,.
\]
We see that the dissipation coefficient for blue detuning ($\Delta>0$)
is always negative, giving rise to runaway solutions. Therefore the
solutions near the north pole are always unstable under blue detuning,
independent of the light intensity. These trajectories run to a fixed
point near the south pole, which accepts stable solutions for $\Delta>0$
(switching) or to a limit cycle. Near the south pole, $S_{z}=-S$,
$S_{x}^{0}=SB_{{\rm opt}}/\Omega$ and 
\[
\ddot{\delta S}_{x}=-\Omega\left(\Omega-S\frac{\partial B_{{\rm opt}}}{\partial S_{x}}\right)\delta S_{x}-2GS\kappa\Omega B_{{\rm opt}}\frac{(\Delta-GSB_{{\rm opt}}/\Omega)}{\left[\left(\kappa/\Omega\right)^{2}+(\Delta-GSB_{{\rm opt}}/\Omega)^{2}\right]^{2}}\dot{\delta S}_{x}\,.
\]

Therefore for $\Delta>GSB_{{\rm opt}}/\Omega$ there are stable fixed
points, while in the opposite case there are also runaway solutions
that are caught in a limit cycle. For red detuning, $\Delta\rightarrow-\Delta$
and the roles of south and north pole are interchanged.

\begin{figure}[b]
 
\includegraphics[clip,width=1\columnwidth]{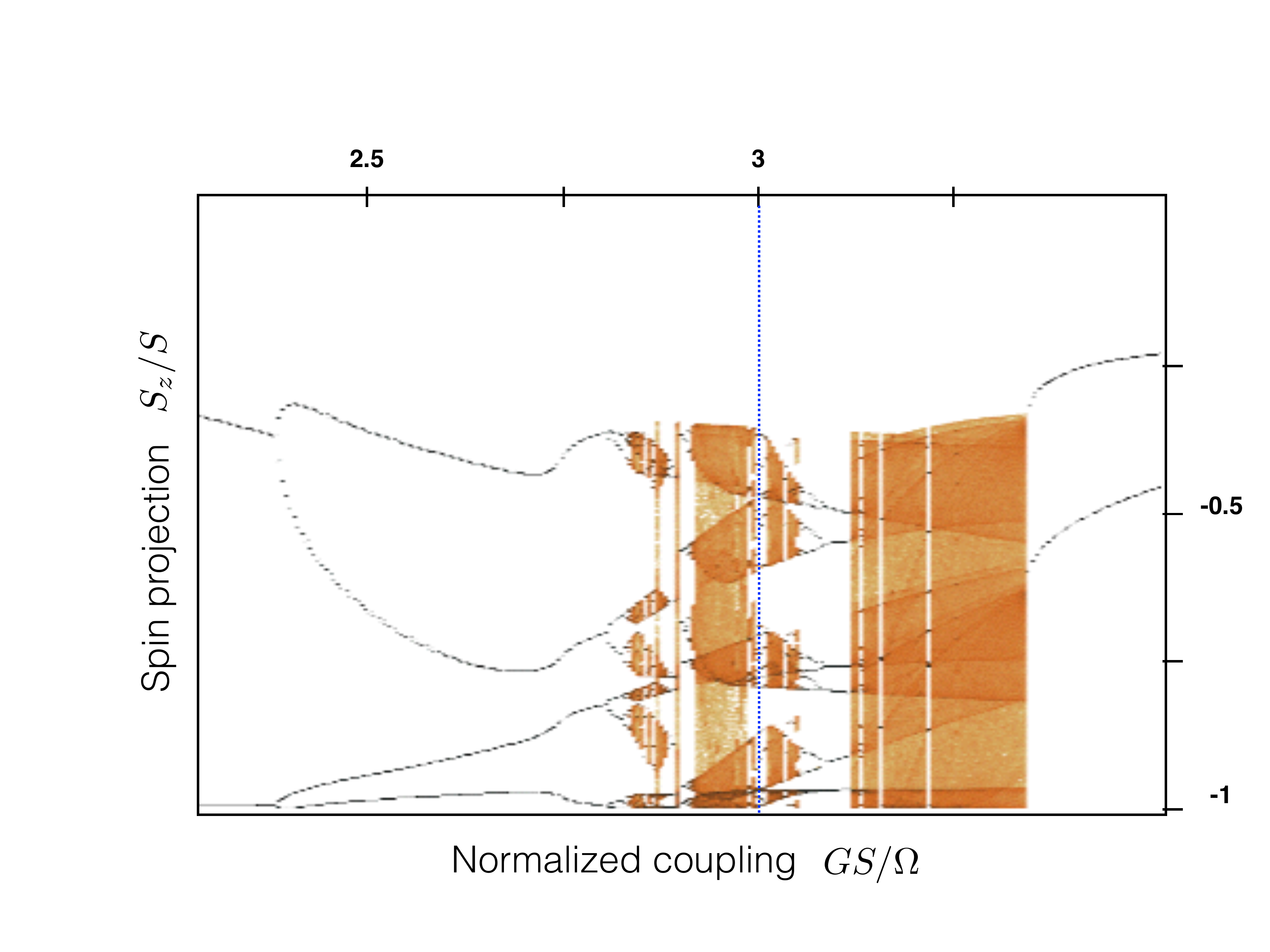}

\caption{(Color online) Bifurcation density plot for $G\alpha_{{\rm max}}^{2}/\Omega=1$
and $\kappa/\Omega=1$ at $\Delta=\Omega$ ($\eta_{{\rm G}}=0$),
as a function of the relative coupling strength $GS/\Omega$. The
dotted blue line indicates $GS/\Omega=3$, for comparison with Fig.
\eqref{Bif}. As in the main text, the points (obtained after the
transient) are given by plotting the values of $S_{z}$ attained whenever
the trajectory fulfills the turning point condition $\dot{S}{}_{z}=0$,
for 20 different random initial conditions. }

\label{Fig:BifSpin} 
\end{figure}

\begin{figure}
\includegraphics[clip,width=1\columnwidth]{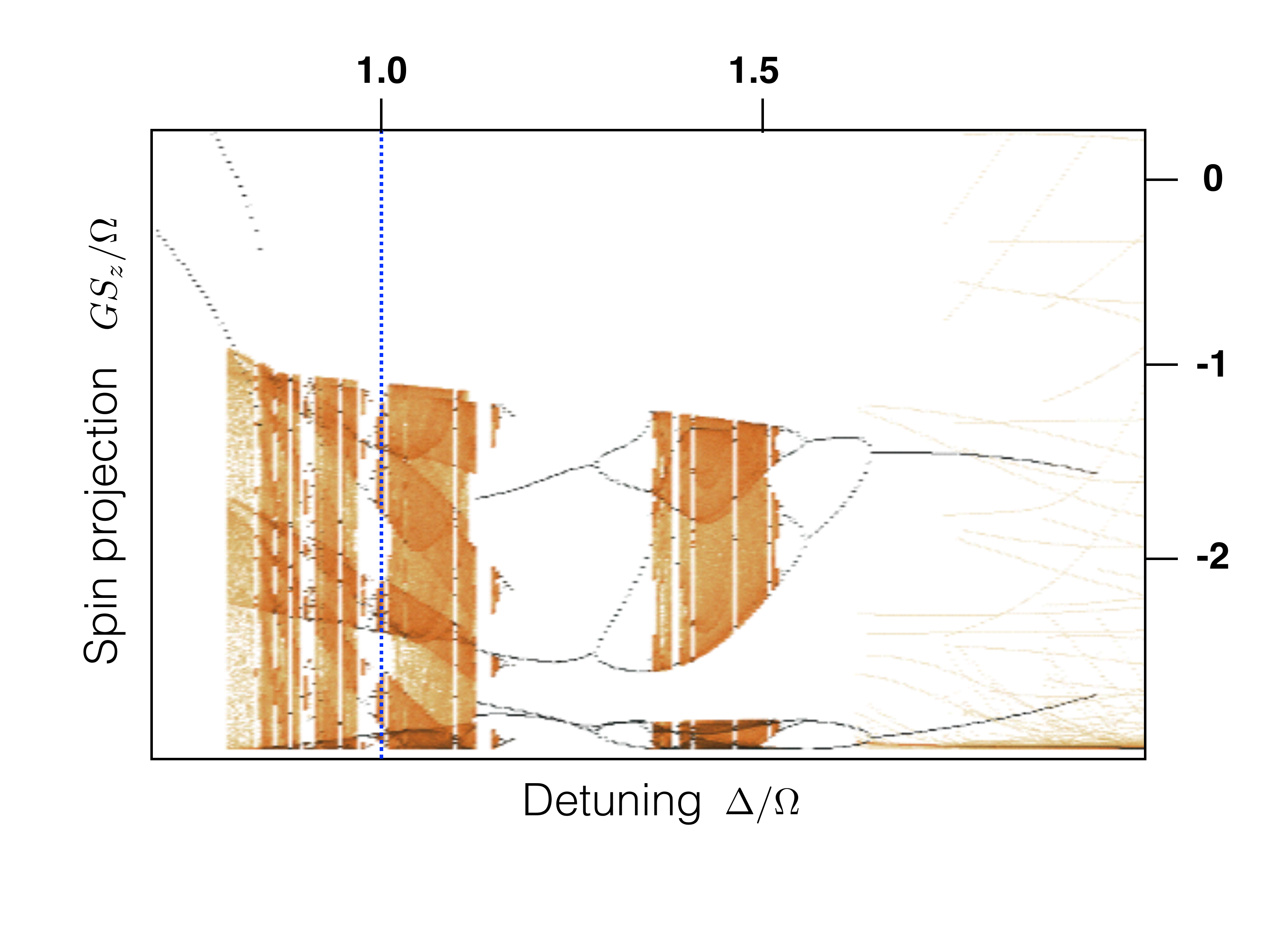}

\caption{(Color online) Bifurcation density plot for $GS/\Omega=3$, $G\alpha_{{\rm max}}^{2}/\Omega=1$
and $\kappa/\Omega=1$ ($\eta_{{\rm G}}=0$), as a function of the
detuning $\Delta/\Omega$. The dotted blue line indicates $\Delta/\Omega=1$,
for comparison with Fig. \eqref{Bif}. }
\label{Fig:BifDelta} 
\end{figure}

\section{Nonlinear dynamics}

\label{sub:SuppNonlinear}

In this section we give more details on the full nonlinear dynamics
described in the main text. In Figs.~\ref{FullDynFig} and \eqref{Bif}
of the main text we chose a relative coupling $GS/\Omega=3$, around
which a chaotic attractor is found. With our estimated $GS\approx10^{10}{\rm Hz}$
for YIG, this implies a precession frequency $\Omega\approx3\cdot10^{9}{\rm Hz}$.
In Fig.~\eqref{FullDynFig} the chaotic regime is reached at $\kappa\approx\Omega/2$
with $G\alpha_{{\rm max}}^{2}/\Omega=1$, which implies $\alpha_{{\rm max}}^{2}\approx S/3$,
that is, a number of photons circulating in the (unperturbed) cavity
of the order of the number of locked spins and hence scaling with
the cavity volume. Bigger values of the cavity decay rate are allowed
for attaining chaos at the same frequency, at the expense of more
photons in the cavity, as can be deduced from Fig. \eqref{Bif} where
we took $\kappa=\Omega$. On the other hand we can think of varying
the precession frequency $\Omega$ by an applied external magnetic
field and explore the nonlinearities by tuning $GS/\Omega$ in this
way (note that $GS$ is a material constant). This is done in Fig.
\eqref{Fig:BifSpin}. Alternatively, the nonlinear behavior can be
controlled by varying the detuning $\Delta$, as shown in Fig. \eqref{Fig:BifDelta}.

\section{Relation to the optomechanical attractors}

\label{sub:SuppOptomech}

In this appendix we show that the optomagnonic system includes the
higher order nonlinear attractors found in optomechanics as a subset
in parameter space.
\begin{figure}
\includegraphics[width=12cm]{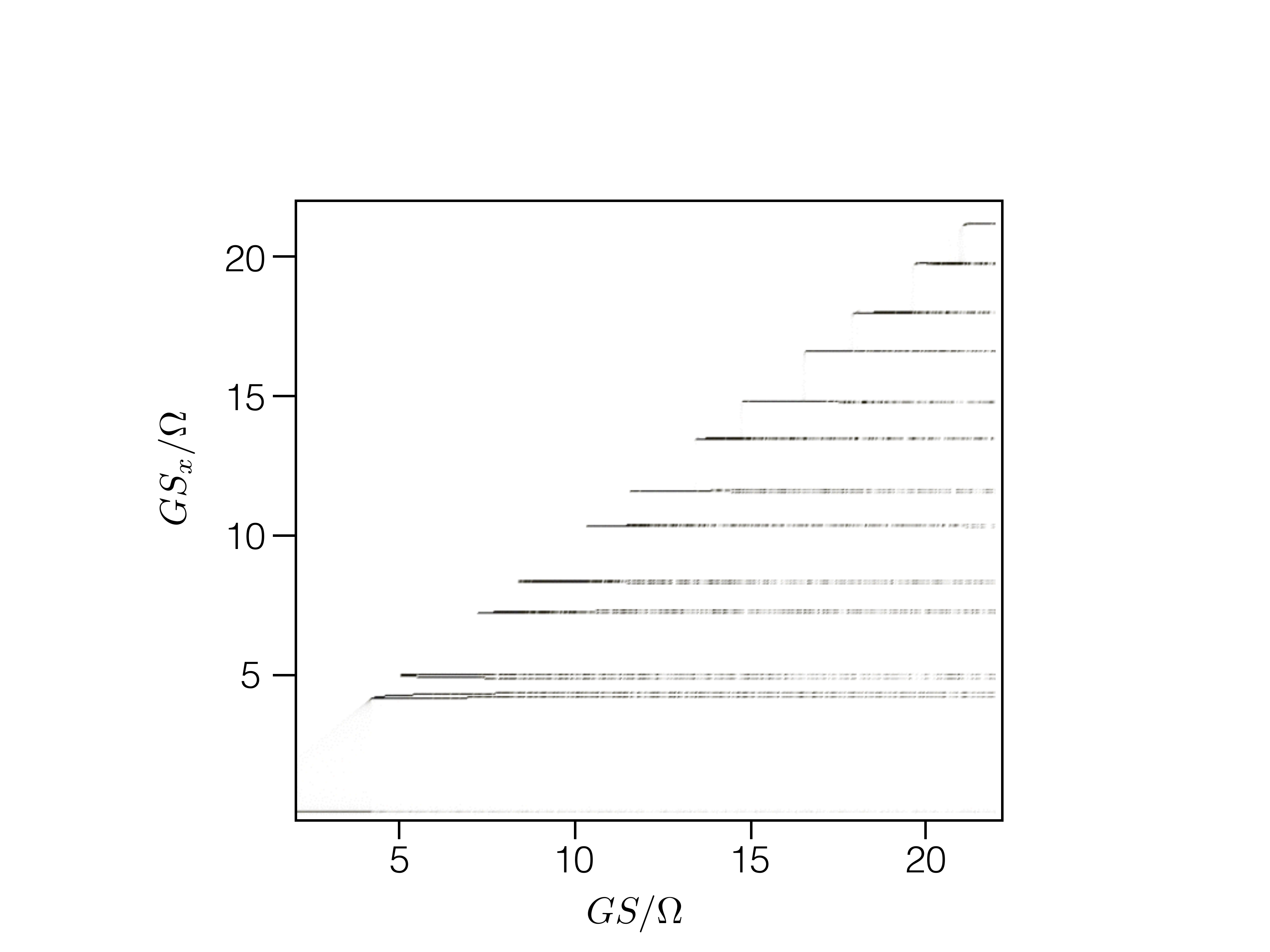}

\includegraphics[width=14cm]{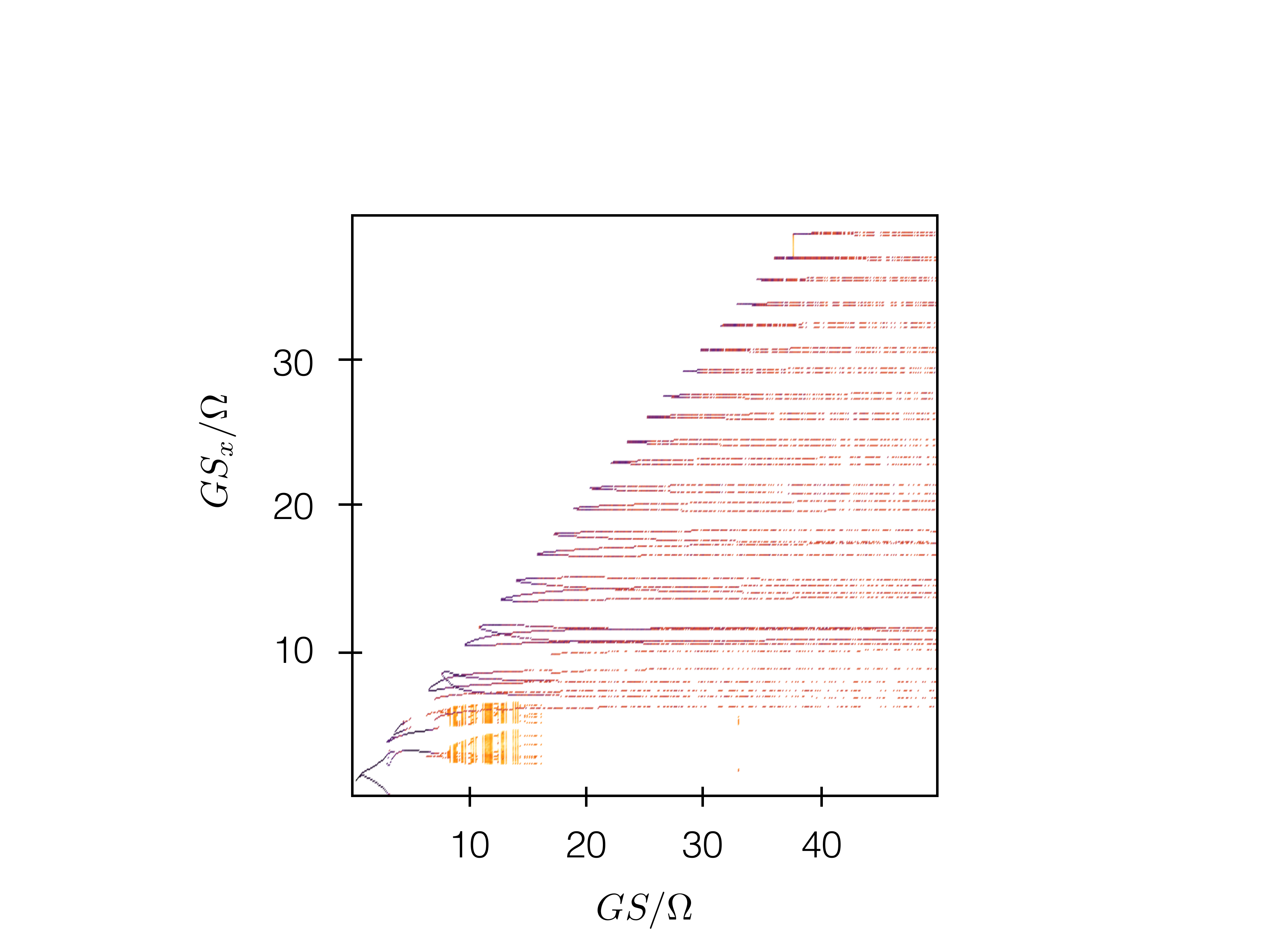}\caption{Attractor diagram for $\Delta=1.5\Omega$ and $\kappa/\Omega=1$ with
condition $G^{2}S|\alpha_{{\rm max}}|^{2}=n\Omega^{2}$. Top: $n=1$,
bottom $n=10$. We plot the $S_{x}$ values attained at the turning
points ($\dot{S}_{x}=0$) for $S_{x}>0$. The diagram is symmetric
for $S_{x}<0$ as expected for a limit cycle on the Bloch sphere.
The diagram at the left coincides to a high degree of approximation
with the predictions obtained for optomechanical systems (i.e. replacing
the spin by a harmonic oscillator). In contrast, this is no longer
the case for the diagram on the right, which involves higher light
intensities.}

\label{Fig:BifOpto} 
\end{figure}

In optomechanics, the high order nonlinear attractors are self sustained
oscillations with amplitudes $A$ such that the optomechanical frequency
shift $GA$ is a multiple of the mechanical frequency $\Omega$. Translating
to our case, this means $G\delta S\sim n\Omega$. Since $\delta S/S\sim G|\alpha_{{\rm max}}|^{2}/\Omega=B_{\alpha_{{\rm max}}}/\Omega$
we obtain the condition 
\begin{equation}
\frac{GS}{\Omega}\frac{B_{\alpha_{{\rm max}}}}{\Omega}\sim n\,\label{eq:OptCond}
\end{equation}
for observing these attractors. We can vary $B_{\alpha_{{\rm max}}}$
according to Eq. \eqref{eq:OptCond}. For $\Omega/GS\ll1$ we are
in the limit of small $B_{\alpha_{{\rm max}}}/\Omega$ and we expect
limit cycles precessing along ${\bf e_{z}}$ as discussed in Sec.
\eqref{sec:Discussion}. In Fig. \ref{Fig:BifOpto} the attractor
diagram obtained by imposing condition \eqref{eq:OptCond} is plotted.
Since the trajectories are in the $xy$ plane, we plot the inflection
point of the coordinate $S_{x}$. We expect $GS_{x}/\Omega$ evaluated
at the inflection point, which gives the amplitude of the limit cycle,
to coincide with the optomechanic attractors for small $B_{\alpha_{{\rm max}}}/\Omega$
and hence flat lines at the expected amplitudes (as calculated in
Ref. \cite{Marquardt2006}) as $GS/\Omega$ increases. Relative evenly
spaced limit cycles increasing in number as larger values of $GS/\Omega$
are considered are observed, in agreement with Ref. \cite{Marquardt2006}.
Remarkable, these limit cycles attractors are found on the whole Bloch
sphere, and not only near the north pole where the harmonic approximation
is strictly valid. These attractors are reached by allowing initial
conditions on the whole Bloch sphere. For $n=1$, (Fig. \ref{Fig:BifOpto},
top), switching is observed up to $GS/\Omega\sim4$ and then perfect
optomechanic behavior. For higher values of $n$, deviations from
the optomechanical behavior are observed for small $GS/\Omega$ (implying
large $B_{\alpha_{{\rm max}}}/\Omega$ according to Eq. \eqref{eq:OptCond})
and large amplitude limit cycles, as compared to the size of the Bloch
sphere. An example is shown in Fig. \ref{Fig:BifOpto}, bottom, for
$n=10$.
\end{widetext}

\end{document}